\documentclass[pra,aps,amsfonts,amsmath,superscriptaddress,%
twocolumn,showpacs,floatfix]{revtex4}
\usepackage{graphicx, latexsym, amssymb, amsmath, color, multirow}
\usepackage{epsfig}
\usepackage{dcolumn}
\usepackage{bm}
\usepackage{eucal}

\def\3j#1#2#3#4#5#6{\left({{#1}\atop{#4}}\quad{{#2}\atop{#5}}
 \quad{{#3}\atop{#6}}\right)}
\def\6j#1#2#3#4#5#6{\left\{{{#1}\atop{#4}}\quad{{#2}\atop{#5}}
 \quad{{#3}\atop{#6}}\right\}}
\def\9j#1#2#3#4#5#6#7#8#9{\left\{\begin{array}{ccc} #1&#2&#3\\
#4&#5&#6\\ #7&#8&#9 \end{array}\right\}}

\def\bm#1{\mbox{\boldmath $#1$}}

\begin{document}
\title{Wigner-Seitz cells in neutron star crust with finite range interactions }
\author{Hoang Sy Than}
\affiliation{Institut de Physique Nucl\'eaire, Universit\'e Paris-Sud, IN2P3-CNRS, F-91406 Orsay Cedex, France}
\affiliation{Institute for Nuclear Science and Technique, VAEC, 179
Hoang Quoc Viet Road, Nghia Do, Hanoi, Vietnam}
\affiliation{Vietnam Agency for  Radiation and Nuclear Safety (VARANS),
70 Tran Hung Dao, Hanoi, Vietnam}
\author{E. Khan}
\affiliation{Institut de Physique Nucl\'eaire, Universit\'e Paris-Sud, IN2P3-CNRS, F-91406 Orsay Cedex, France}
\author{Nguyen Van Giai}
\affiliation{Institut de Physique Nucl\'eaire, Universit\'e Paris-Sud, IN2P3-CNRS, F-91406 Orsay Cedex, France}

\date{\today}

\begin{abstract}
The structure of Wigner-Seitz cells in the inner crust of neutron stars is
investigated using a microcospic Hartree-Fock-BCS approach with finite
range D1S and M3Y-P4 interactions. Large effects on the densities are
found compared to previous predictions using Skyrme interactions. Pairing
effects are found to be small, and they are attenuated by the use of finite range
interactions in the mean field.
\end{abstract}
 \maketitle

\section{Introduction}

In recent years, the properties of the inner crust of neutrons stars have
been investigated within various frameworks, especially focusing on their
microscopic structure and superfluid properties. The microscopic
calculations of the inner crust matter are usually studied in the
Wigner-Seitz (WS) approximation \cite{Ne73,Pe95}. Following the standard
approaches, the inner crust consists of a lattice of WS cells, each cell
containing a neutron-rich nucleus immersed in a sea of dilute gas of
neutrons and relativistic electrons uniformly distributed inside the cell
\cite{Pe95}.The first microscopic calculations of the properties of the
inner crust matter were done by Negele and Vautherin in the 70s \cite{Ne73}.
The calculations were performed assuming a set of non-interacting cells
described in the Hartree-Fock (HF) approach with 11 representative cells.
These cells are distributed in different zones of the inner crust with
densities covering a range from 1.743 $\times 10^{-3} \rho_0$ to
0.5$\rho_0$, $\rho_0$ = 0.16 fm$^{-3}$ being the nuclear matter saturation
density. The optimal number ($N,Z$) of neutrons and protons in each cell is
obtained by searching for the lowest binding energy satisfying the
$\beta$-stability condition of the cell.

More recently, the superfluid properties and their influence on the specific heat
were investigated in the self-consistent Hartree-Fock-Bogoliubov (HFB)
approach \cite{Ni04, Ni04b}. The collective excitations and the cooling time
of the inner crust of neutron stars were also studied in the framework of a
spherically symmetric HFB + quasiparticle random phase approximation \cite{El05, Gr08b}
and of HFB approach at finite temperature \cite{Mo07}.

All the above calculations were done within the non-relativistic framework
with the SLy4 \cite{Ch98} Skyrme
interaction in the mean field channel and a density-dependent delta force
for the pairing interaction. Later on,
the properties of the WS cells have been studied in the non-relativistic
work of Baldo {\it et al.} \cite{Bal05, Bal05N, Bal06} using an energy
functional method including the pairing correlations of protons and neutrons. The
relativistic Hartree, or
Relativistic Mean Field (RMF) approach \cite{Cao08} has also been applied to
investigate the structure of WS cells and the influence of different boundary conditions on
the structure of the cells.

However, 
there is so far no study of the inner crust in the HF approach using
finite range interactions.
The aim of this paper is to examine the
behaviour of density distributions of protons and neutrons in the different WS cells
obtained with a finite range interaction within a self-consistent HF-BCS
calculation, where the HF equations are solved with the Dirichlet-Neumann
mixed boundary conditions. For the numerical studies one obvious choice is
the D1S interaction \cite{Ro87} which is widely employed in finite nuclei calculations. More
recently, the M3Y-P4 interaction proposed by Nakada \cite{Na09} seems to work
satisfactorily well for describing nuclear ground states in the HFB framework.
This finite-range effective interaction derives from the well-known M3Y interaction
and it is interesting to study its predictions for neutron stars.

The structure of this paper is as follows. In Section II we present the HF
and HF-BCS approaches 
using finite-range
density-dependent interactions in both the mean field and pairing channels.
The results of the calculations are presented in Section III.
The evolution of the neutron and proton densities as well as the neutron pairing fields
in the regions of the inner crust are discussed.
The summary and perspectives of this study are given in Section IV.

\section{HF-BCS calculations with finite range interactions}


A practical difficulty of HF calculations in WS cells is that the cell
radius can be large (up to 40-50 fm). Solving directly the
integro-differential equations in coordinate space can lead to inaccuracies
and instabilities \cite{Tha09}. The method that we adopt consists in solving the HF
equations in a basis well adapted to each WS cell under study. In subsection
A we derive the HF equations in coordinate space, and in subsection B we
give some details on their solution.

\subsection{HF equations in coordinate space}
We summarize here the analytical expressions needed for HF calculations with
a finite-range interaction, taking as an illustrative example the Gogny
interaction \cite{Dec80,Ro87} which contains a sum of two Gaussians, a
zero-range density-dependent part and a zero-range spin-orbit part.
Expressions for other types of finite-range interactions can be easily
deduced from this case. For example, the M3Y-Pn interactions of Nakada 
\cite{Na09} have zero-range, density-dependent terms just like Gogny and
Skyrme interactions and therefore, the corresponding contributions to the
mean field potentials can be deduced from the expressions given here. In
addition, the M3Y-Pn interactions also contain tensor terms, except for the
M3Y-P4 interaction that we use in this work and therefore, we do not discuss
the tensor contributions. Finally, the original M3Y-P4 interaction contains
short-range spin-orbit components (with $\mu=0.25 fm$ and $0.40 fm$, see Eq.
\ref{th15}) that we approximate for simplicity by a zero-range spin-orbit
component as in the Gogny force. The corresponding spin-orbit strength 
$W_0=160 MeV.fm^5$ is adjusted on the empirical $1p1/2-1p3/2$ proton splitting in
$^{16}$O as in \cite{Dec80}.
Throughout this
work we limit ourselves to spherically symmetric systems, which should apply to baryonic densities in
the inner crust
ranging from $1.4 10^{-3} \rho_0$ to about
$0.5 \rho_0$, where $\rho_0$=$0.16 fm^{-3}$ is the nuclear matter saturation
density \cite{Ne73,Dou00}.

In order to obtain the Hartree and Fock potentials,
 we use the multipole expansion
of the Gaussian form factor \cite{Br93}:
\begin{equation}
e^{\frac{-|\bm{r_1}-\bm{r_2}|^2}{\mu^2_\nu}}=  4\pi\sum_{LM} (-)^M v^{\nu}_L(r_1,r_2)Y^{-M}_L
(\hat{r_1})Y^M_L (\hat{r_2})~. \label{th15}
\end{equation}
We express the central part of the Gogny force in the form
\begin{eqnarray}
V(|\bm{r_1}-\bm{r_2}|)=4\pi\sum_{SLJ}\sum^2_{\nu=1}&&A_{\nu}(S)(-1)^{L+S+J+M}v^{\nu}_L(r_1,r_2) \nonumber \\
&&\left (T^{(SL)J}_{(1)}.T^{(SL)J}_{(2)} \right)~, \label{th16}
\end{eqnarray}
where
\begin{eqnarray}
A_{\nu}(S=0)&=&W_{\nu}-H_{\nu}P^{\tau}+\frac{B_{\nu}-M_{\nu}P^{\tau}}{2}~, \nonumber \\
A_{\nu}(S=1)&=&\frac{B_{\nu}-M_{\nu}P^{\tau}}{2}~. \label{th39}
\end{eqnarray}
Here, $W_{\nu}$, $B_{\nu}$, $H_{\nu}$, $M_{\nu}$, $\mu_{\nu}$ are the parameters of the Gogny
interaction or appropriate combinations of the parameters of the M3Y-Pn interactions, $P^\sigma$
and $P^\tau$ are the spin and
isospin exchange operators,
respectively. We have introduced the tensors $T^{(SL)J}_{(\mu)}$ which are tensorial products of a
 spherical harmonic $Y^M_L$ with a Pauli spin matrix
\begin{equation}
T^{(SL)J} = \left [\sigma^S_\eta \otimes Y^M_L \right].
\label{th17}
\end{equation}

Because of the spherical symmetry assumption, the single-particle
wave functions $\varphi_i(\bm{r},\sigma,q)$ can be
factorized into a radial part $u_i(r)$, a spin-angular part
 $\mathcal{Y}_{ljm}(\hat{r},\sigma)$, and an isospin
part $\chi_{q}(\tau)$:
\begin{equation}
\varphi_i(\bm{r},\sigma,q)=\frac{u_\alpha(r)}{r}
\mathcal{Y}_{ljm}(\hat{r},\sigma)\chi_{q}(\tau) \label{th19*}
\end{equation}
where
\begin{equation}
 \mathcal{Y}_{ljm}(\hat{r},\sigma) \equiv\sum_{m_lm_s}<l\frac{1}{2}m_lm_s|jm>
Y_{lm_l}(\hat{r})\chi_{m_s}(\sigma)~,
\end{equation}
 $\chi_{m_s}(\sigma)$ being a spinor corresponding to a spin projection $m_s$, and the index $i$ stands for the following set of quantum
numbers: the charge $q$, the principal quantum number $n$, the
orbital angular momentum $l$, the total angular momentum $j$, and
the magnetic quantum number $m$.


Using the tensors $T^{(SL)J}_{(\mu)}$ 
and the
single-particle wave functions $\varphi_i(\bm{r},\sigma,q)$
we can calculate the direct (Hartree) and
exchange (Fock) potentials 
in each ($l,j$) partial wave. 
With the help of the results (\ref{th26}-
\ref{th31}) of Appendix A  it is
straightforward to obtain the radial integro-differential HF equations in coordinate space
 for the radial wave functions $u_i(r_1)$:
 \begin{widetext}
\begin{eqnarray}
\frac{\hbar^2}{2m}[-u^{''}_i(r_1)&+&\frac{l_i(l_i+1)}{r^2_1}u_i(r_1)]
+  U^D_i(r_1)u_i(r_1)- \int U^E_i(r_1,r_
2)u_i(r_2)r^2_2 dr_2 \nonumber \\ &+&
[j_i(j_i+1) -l_i(l_i+1)-\frac{3}{4}
]W^{LS0}_q(r_1)u_i(r_1)=\epsilon_iu_i(r_1)
 \label{th32}
\end{eqnarray}
\end{widetext}
where the local central potential $U^D_i$ contains the direct contributions of the finite range forces
(nuclear and Coulomb) as well as the direct+exchange contributions of the zero-range
density-dependent forces,  while the direct+exchange contributions of the zero-range
spin-orbit force are in both the $W^{LS1}_q$ component of $U_i^D(r)$
and in the one-body spin-orbit potential $W^{LS0}_q$\cite{Ch98}:  
\begin{equation}
U^D_i(r_1)=U^H_i(r_1)+U^{DD}_q(r_1)+V^{DC}(r_1)+W^{LS1}_q(r_1)~.
\end{equation}
The non-local potential $U^E_i$
is composed of the exchange contributions of all finite range (nuclear and Coulomb) forces: 
\begin{equation}
U^E(r_1,r_2)=U^F_i(r_1,r_2)+V^{EC}_i(r_1,r_2)~.
\end{equation}
The corresponding expressions are given in Appendix A.

\subsection{Solving HF equations in a basis representation}
Solving the HF equations directly in coordinate space has several advantages: the
results do not depend on the choice of a basis and its truncation, and the
individual wave functions have a correct asymptotic behavior. However, 
in the regions of the inner crust where the density is below $10^{-2}\rho_0$ the
radius of the WS cells can be larger than $40$ fm and strong numerical instabilities may appear
for single-particle wave functions with large orbital momentum ($l\sim 15-20$). The
cause of the problem is that the HF wave functions behave like
$r^{l+1}$ 
near the origin. Calculations requiring a
certain number of nodes can fail 
if the HF potential is not
accurate enough at a given iteration. To avoid this dificulty we have
developed a basis expansion method \cite{Tha09} for solving 
the HF equations as a matrix diagonalization problem. The advantage of
the basis expansion method is that it produces a high accuracy (measured by
the orthogonality of the solutions) with a smaller number of points, while
this may not be obtained with 
the coordinate space method. 

The harmonic-oscillator functions basis is the most common choice for expanding the HF
single-particle wave functions.
Here, we choose instead, as a basis,
the spherical Bessel functions normalized inside the Wigner-Seitz cell sphere.

The radial part $\frac{u_{nlj}(r)}{r}$ of the single-particle wave
function in Eq.~(\ref{th19*}) can be expanded on the normalized
spherical Bessel functions 
as
\begin{equation}
\frac{u_{nlj}(r)}{r}=\sum^N_{i=1} C_{nlj,i}
\tilde{j_l}(k^{(l)}_ir) \label{th46}
\end{equation}
where $N$ is the dimension of the basis and $\tilde{j_l}(k^{(l)}_ir)$
the normalized spherical Bessel functions with boundary conditions that will be specified later.
The single-particle wave function $\varphi_i(\bm{r},\sigma,q)$ of
Eq.~(\ref{th19*}) can be expanded as
\begin{equation}
\varphi_{nljm}(\bm{r},\sigma,q)=
\sum^N_{i=1} C_{nlj,i} \psi_{nljm,i}(\bm{r},\sigma,q)~,
\label{th48}
\end{equation}
where the orthonormal basis is
\begin{equation}
\psi_{nljm,i}(\bm{r},\sigma,q)=\tilde{j_l}(k^{(l)}_ir)\mathcal{Y}_{ljm}(\hat{r},\sigma)\chi_{q}(\tau)~.
\label{new1}
\end{equation}
Under the $l$ and $j$ conservation the HF
Hamiltonian matrix is decomposed into $(l,j)$ blocks:
\begin{equation}
H^{(lj)}_{ii'}=<\psi_{ljm,i}|H|\psi_{ljm,i'}>~, \label{th49}
\end{equation}
where $H$ stands for the HF Hamiltonian.
The HF equations then become:
\begin{equation}
\sum^N_{i'=1}H^{(lj)}_{ii'}C_{nlj,i'}=\epsilon_{nljm}
C_{nlj,i} \label{th50}
\end{equation}
where $H^{(lj)}_{ii'}$ is a symmetric $N\times N$ matrix,
\begin{eqnarray}
H^{(lj)}_{ii'}&&=\frac{\hbar^2}{2m}(k^{(l)}_i)^2\delta_{ii'}+ \\
&&\int\int
V^{HF}_{lj}(r_1,r_2) \tilde{j_l}(k^{(l)}_ir_1)
\tilde{j_l}(k^{(l)}_{i'}r_2) d\bm{r_1}d\bm{r_2}~. \nonumber
\label{th51}
\end{eqnarray}
Here, the potentials $V^{HF}_{lj}(r_1,r_2)$ 
can be obtained
from the formulas (\ref{th26} - \ref{th31}) of Appendix A.
The HF equations are solved by an iterative
procedure, starting from an initial Woods-Saxon potential. 
In the present calculation we use a basis size N=20.
The convergence criteria is set
on the single-particle energies $\epsilon_{nljm}$ to be 1 keV. 

The HF calculations are done with finite-range interactions (the D1S
\cite{Ro87} and M3Y-P4 \cite{Na09} forces) imposing Dirichlet-Neumann
boundary conditions at the edge of the cell as introduced in Ref.
\cite{Ne73}. These boundary conditions for the single-particle wave
functions are taken as follows: (i) the even parity wave functions vanish at
the edge $r$ = $R_{WS}$ of the box; (ii) the first derivative of the
odd-parity wave functions vanish at $r$ = $R_{WS}$. The
purpose of these chosen boundary conditions 
is to obtain an approximately constant density at large distance 
from the center of the cell, thus simulating a lattice of nucleus-like systems
embedded in a uniform neutron gas.

Of course, these boundary conditions are somewhat arbitrary 
and one could as well choose alternative boundary conditions.
Thus, another kind of boundary conditions could be chosen in the following
way: odd $l$ wave functions vanish at $r$ = $R_{WS}$, the first derivatives of even $l$
wave functions vanish at $r$ = $R_{WS}$.
As shown in Ref. \cite{Bal06}, the two kinds of boundary conditions can
be used in the calculations of the neutron star inner crust. The difference
of the binding energies per nucleon for each cell will increase with the
increasing density (see Table 1 of Ref. \cite{Bal06}). However, the values
of these uncertainties are smaller than the variations of the equilibrium
configuration connected with the pairing effects \cite{Bal05, Bal06}.
Therefore, in our study the boundary conditions and the spherical WS cell
structure are kept the same as in Ref. \cite{Ne73}. We do not redetermine
the number ($N,Z$) of neutrons and protons in the considered cells. This study
can be envisaged in the near future.

It should be noted that the Bloch boundary conditions can be used at the cost
of more complex calculations 
as presented in Refs.
\cite{Cha07,Cha07b,Has08}. The validity in neutron star crust
of the Wigner-Seitz approximation that we use here has been
discussed in \cite{Cha07}.

\subsection{Pairing correlations in BCS approximation}

The effects of pairing correlations are known to be substantial in the inner
crust \cite{Ni04,Ni04b} and it is necessary to include them in order to
obtain a more realistic picture. In this first study of the inner crust with
finite range interactions in the particle-hole (mean field) channel we adopt
the simplified BCS picture for describing the neutron pairing fields, but
the HF-BCS model remains completely self-consistent. Indeed, the BCS
occupation numbers are recalculated at each HF iteration.

Various effective interactions can be used in the pairing channel. In this
work the pairing correlations are treated in the BCS approximation with both
the zero-range density-dependent and finite range density-dependent forces
in the pairing channel.

The importance of the density dependence of the pairing interaction is well
known in the theories of superfluidity in neutron stars. As shown in Ref.
\cite{Pe95}, it is impossible to deduce the magnitude of the pairing gaps in
neutron stars with sufficient accuracy. The calculation of the $^1S_0$
pairing gaps in pure neutron matter, or symmetric nuclear matter based on
bare {\it NN} interaction depends strongly on the forces that are used.

In this work, we compare the results of the zero-range 
and finite-range
interactions as pairing interactions. For the finite-range interactions, the
effective Gogny D1S 
\cite{Ro87} or M3Y-P4 \cite{Na09} interactions
 are used to calculate the pairing field. 

For the zero-range force in the pairing channel, we use the
form
\begin{equation}
V(\bm{r}_1-\bm{r}_2) = V_0\left ( 1-\eta\left ( \frac{\rho(\bm{r})}{\rho_0} \right
)^{\alpha}\right ) \delta(\bm{r}_1-\bm{r}_2)
\label{th73}
\end{equation}
where $V_0$, 
$\eta$ and $\alpha$ are parameters that can be adjusted. 
We have studied the predictions of
two types of pairing forces: a 
density-independent
interaction ($\eta=0$) that gives rise to volume pairing and a
density-dependent delta force that can give rise to surface pairing.
In Appendix B, we show the details of
the calculations of pairing matrix elements.

Up to now, the magnitude of pairing correlations in neutron matter is still
a subject of debate. 
The D1S Gogny interaction commonly used in finite nuclei
calculations gives a maximum value of pairing gap in infinite neutron
matter of about 2.4 MeV at a Fermi momentum $k_F \approx $
0.8 fm$^{-1}$ \cite{Ga99}. The maximum value of pairing gap in infinite
matter using the M3Y-P4 interaction is about 3.2 MeV at the similar Fermi
momentum \cite{Na09}. On the other hand, the microscopic calculations of
Refs. \cite{Wa93,Sch03} predict for the maximum gap a value of about 1
MeV. One observes that there are three different scenarios for pairing
correlations in neutron matter. In our study, we use the density dependent
delta force for the pairing interaction to simulate the third scenario. 
This is obtained by adopting the parameter values of Ref.\cite{Ni04}:
$V_0$=-330 MeV fm$^{-3}$, $\eta$=0.7, and
$\alpha$=0.45 .
Thus, for each WS cell we 
perform three HF-BCS calculations with three different pairing forces. The BCS pairing
window is chosen to be $\pm$ 6 MeV around the Fermi level.

\section{Results and Discussions}

In the present study, we have performed the HF-BCS calculations for a set of
11 representative WS cells determined in Ref. \cite{Ne73}. The considered
density range is from neutron drip density $\rho_{min}$ = 1.743 $\times$
10$^{-3}$ $\rho_0$ to about $\rho_{max}$ = 0.5$\rho_0$. In this density
range, the nuclear clusters are assumed to be spherical \cite{Dou00}. Above
the density $\rho_{max}$ the energy per baryon approaches the value of the
uniform neutron system and the cells in the inner crust might deviate from
the spherical shape \cite{Mag04}. Following Ref. \cite{Ne73} we denote the
WS cells like a nucleus with $Z$ protons and $N$ neutrons. The eleven zones
of the representative cells of the inner crust with mean densities and
corresponding proton number $Z$ and neutron number $N$ in each cell are
listed in Table \ref{t5}. The decreasing zone number $N_{zone}$ are from 10
to 0, corresponding to the increasing density from the minimum density
$\rho_{min} = 2.79 \times 10^{-4}$ fm$^{-3}$ to the maximum density
$\rho_{max} = 7.89 \times 10^{-2}$ fm$^{-3}$. The WS cells are denoted like
a nucleus as $^{180}$Zr, $^{200}$Zr, $^{250}$Zr, $^{320}$Zr, $^{500}$Zr,
$^{950}$Sn, $^{1100}$Sn, $^{1800}$Sn, $^{1350}$Sn, $^{1500}$Zr, and
$^{982}$Ge, as introduced in Ref. \cite{Ne73}. These (N,Z) values must be
considered as indicative since they were determined by using a different
model for the HF mean field in Ref.\cite{Ne73}. The WS cell radii $R_{WS}$
are calculated by the following relation
\begin{equation}
<\rho>=\frac{A}{\frac{4\pi}{3}R^3_{WS}}
\end{equation}
where $\rho$ and $A$ are the density and mass number of the
considered cell, respectively. The values of the radii $R_{WS}$
are also shown in the last column of Table \ref{t5}.

\begin{table}
\caption{The Wigner-Seitz cells considered in this work. $\rho$,
$N$, $Z$ and $R_{WS}$ are the baryonic densities, the numbers of
neutrons and protons, and the WS cell radii,
respectively. All values are taken from Ref.
\cite{Ne73}. } {\begin{tabular}{| c| c| c| c| c|} \hline \hline
 $N_{zone}$ & $\rho/\rho_0$ & N & Z & $R_{WS} [fm]$  \\ \hline
10 & 0.143 $\times$ 10$^{-2}$ & 140 & 40 & 53.6 \\
9& 0.250 $\times$ 10$^{-2}$ & 160 & 40 & 49.2 \\
8 & 0.375 $\times$ 10$^{-2}$ & 210 & 40 & 46.3 \\
7 & 0.549 $\times$ 10$^{-2}$ & 280 & 40 & 44.3 \\
6 & 0.994 $\times$ 10$^{-2}$ & 460 & 40 & 42.2 \\
5 & 0.233 $\times$ 10$^{-1}$ & 900 & 50 & 39.3 \\
4 & 0.361 $\times$ 10$^{-1}$ & 1050 & 50 & 35.7 \\
3 & 0.557 $\times$ 10$^{-1}$ & 1300 & 50 & 33.1 \\
2 & 1.275 $\times$ 10$^{-1}$ & 1750 & 50 & 27.6 \\
1 & 2.968 $\times$ 10$^{-1}$ & 1460 & 40 & 19.6 \\
0 & 4.931 $\times$ 10$^{-1}$ & 950 & 32 & 14.4 \\
 \hline \hline
\end{tabular}}
\label{t5}
\end{table}

In isolated finite nuclei, for a given number of protons there is always a
maximum number of bound neutrons. This neutron stability limit defines the
neutron drip line. Since the neutron-rich nuclei are quickly beta decaying,
then the neutron drip line is usually drastically limited in the laboratory.
This is not the case for the neutron-rich systems immersed in the inner
crust of neutron stars. In the case of WS cells, the beta decay is blocked
by the presence of the degenerate electron gas uniformly distributed inside
the cell. Therefore, in this case the nuclei inside the inner crust of
neutron stars can bind more neutrons than the nuclei in the vacuum.
Furthermore, there are many delocalized neutrons forming a uniform neutron
gas and filling the outer region of the WS cells.

\subsection{HF and HF-BCS density distributions}

\begin{figure*}
 \hspace*{0cm}\mbox{\epsfig{file=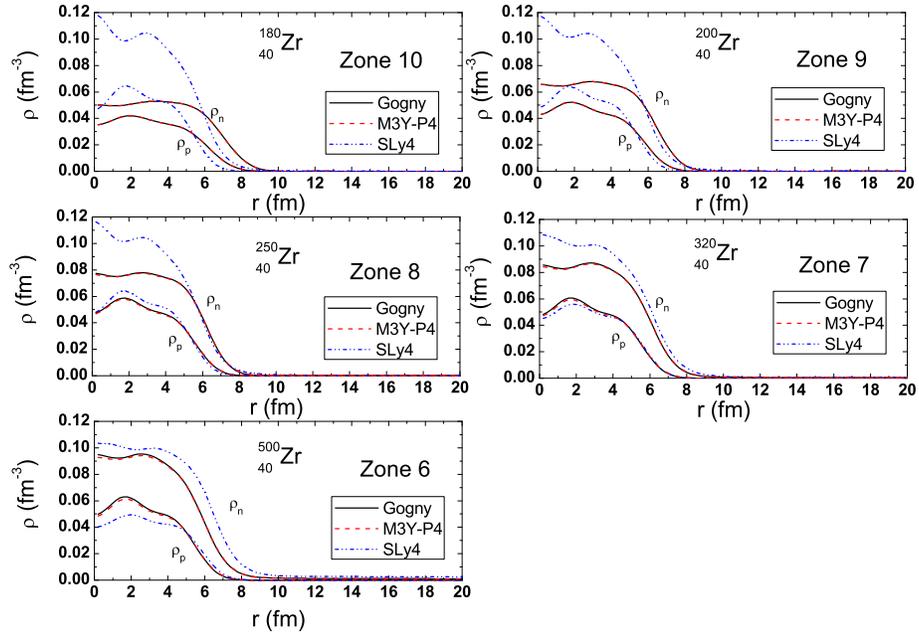,height=10cm}}
\caption{\small (Color online) The HF proton and neutron densities in zones 10 to 6
calculated with the Dirichlet-Neuman boundary conditions and using
D1S \cite{Ro87},
M3Y-P4 \cite{Na09} and SLy4 \cite{Ch98} interactions. 
}
\label{f6}
\end{figure*}

First, we discuss the case of the WS cells calculated in the HF approach
without pairing effects. Fig. \ref{f6} displays the HF proton and neutron
density profiles of $^{180}$Zr, $^{200}$Zr, $^{250}$Zr, $^{320}$Zr and
$^{500}$Zr systems obtained with D1S, M3Y-P4 and SLy4 interactions. One
notes that the numerical HF calculations with SLy4 are the same as in Ref.
\cite{El05}. We observe that the HF calculations with the finite-range D1S
and M3Y-P4 interactions give very similar results, while the results
obtained with SLy4 interaction are different.

We will analyze the case of the cell $^{180}$Zr. One knows that the depopulation
of the $s$ state can lead to a depletion of the central density
\cite{kha08,gra09}. In the case of this cell, the 5$s_{1/2}$ neutron state
is fully filled with SLy4 interaction and empty with D1S and M3Y-P4
interactions. Thus, the neutron densities in the center of the cell obtained
with D1S and M3Y-P4 interactions are smaller by a factor of 2.3 than that
obtained with SLy4 interaction. However, the neutron gas density obtained
with D1S and M3Y-P4 interactions is more than 1.8 times greater than that of
SLy4 interaction. We have checked that if one integrates the neutron density
up to $r$ = 10 fm, where the neutron density profile becomes approximately
constant, one finds about 80 neutrons and 90 neutrons with D1S (or M3Y-P4)
and SLy4 interactions, respectively. Thus, it appears that large surface
regions are observed in the case of D1S and M3Y-P4 interactions, and the
neutron gas density of the cell $^{180}$Zr is much higher. Similar
situations happen in two other cells $^{200}$Zr and $^{250}$Zr. In the case
of cells $^{320}$Zr and $^{500}$Zr, where the 5$s_{1/2}$ neutron state is
fully filled, one observes that the nuclear cluster region becomes larger
with the SLy4 interaction, while its neutron gas density is smaller by a
factor of 2 than those obtained with D1S and M3Y interactions. For the above
cells one concludes that the neutron density in the center of the cell
becomes smaller with D1S and M3Y-P4 interactions.

\begin{figure*}
 \hspace*{0cm}\mbox{\epsfig{file=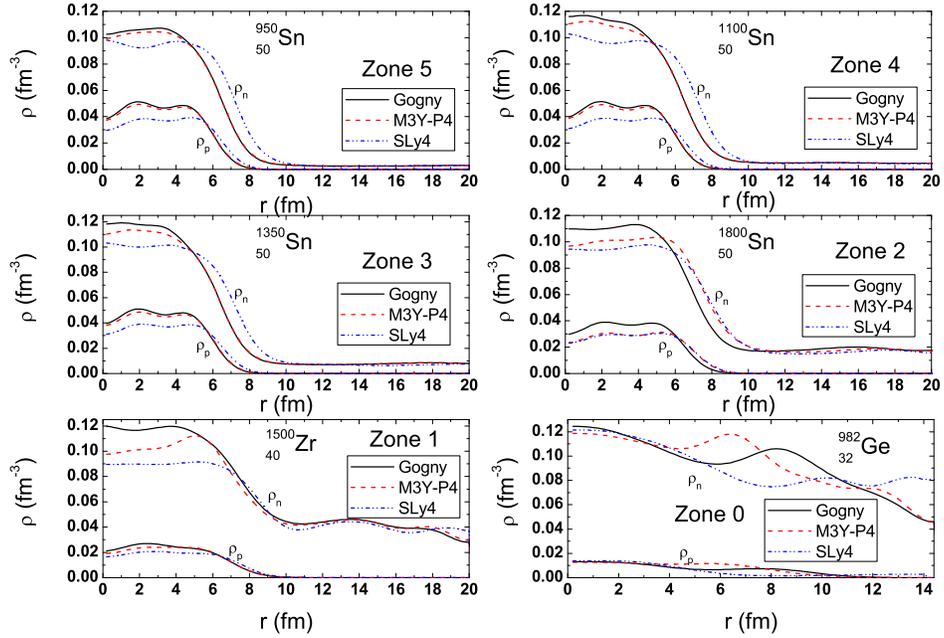,height=10cm}}
\caption{\small (Color online) Same as Fig. \ref{f6}, but for higher density zones (from
zone 5 to zone 0).} \label{f7}
\end{figure*}

However, the situation is opposite at higher
 densities, such as in the cells $^{950}$Sn, $^{1100}$Sn,
$^{1800}$Sn, $^{1350}$Sn and $^{1500}$Zr. These changes can be
seen in Fig. \ref{f7}, where the neutron densities of the nuclear
clusters obtained with SLy4 interaction are always smaller than those
obtained with D1S or M3Y-P4 interactions. One can also see that
the surface thickness of the nuclear cluster with SLy4 interaction becomes larger
by about 10\% in the cells  $^{950}$Sn, $^{1100}$Sn,
$^{1800}$Sn and $^{1350}$Sn. In the cell $^{1500}$Zr, the nuclear cluster surface
 is similar with the three interactions. Since the
neutron density in the center of this cell calculated with SLy4
interaction is smaller than that obtained with D1S or M3Y-P4
interactions, then its outer neutron gas density is a little
higher. For the highest density corresponding to the cell
$^{982}$Ge, there is still some trace of a central cluster and of an outer neutron gas
in the case of SLy4 whereas this separation fades away with D1S and M3Y-P4 interactions.

As mentioned above, the first microscopic calculations of the WS cells in
the inner crust matter were done in Ref. \cite{Ne73}. The pairing effects
were not taken into account in these calculations because it was assumed
that their contributions would be small 
in comparison
with the total binding energy of the considered system. However,
calculations using an energy functional involving the neutron and proton
pairing correlations 
by Baldo {\it et al.} \cite{Bal05N} show that the mean
density $\rho$ of the equilibrium configuration ($Z,
R_{WS}$) can be changed significantly due to the pairing effects. We
therefore calculate also the 
WS cells in HF-BCS approximation. The
D1S and M3Y-P4 interactions are used to construct the mean field and the
pairing field. For comparison we have also considered a hybrid case where a
density-dependent delta force is chosen for the pairing interaction and the
D1S interaction is used in the mean field channel.

\begin{figure*}[htb]
 \hspace*{0cm}\mbox{\epsfig{file=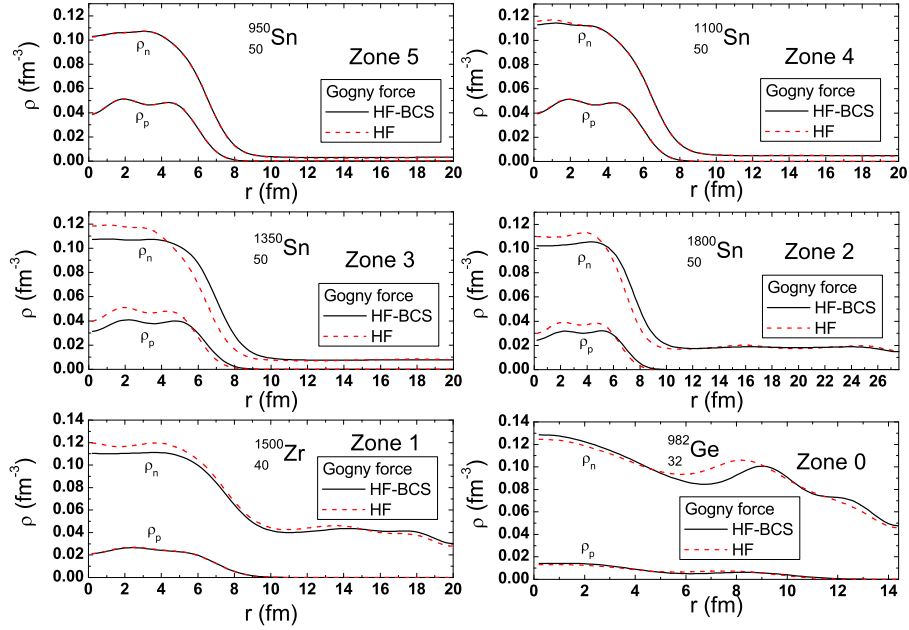,height=10cm}}
\caption{\small \small (Color online) The proton and neutron densities obtained with the
D1S \cite{Ro87} interaction in HF and HF-BCS approximations. The
calculations are done from zone 5 to zone 0 } \label{f21}
\end{figure*}

Fig. \ref{f21} shows the proton and neutron densities
obtained in HF and HF-BCS approximations for 6 WS cells, in which the
D1S Gogny \cite{Ro87} interaction is used in both the mean field
and pairing channels. One can see that the behaviour of the proton
and neutron densities obtained in the two approximations are similar for
$^{950}$Sn and $^{1100}$Sn cells corresponding to the range of
density $\rho \sim$ 2.79 $\times$ 10$^{-4}$ fm$^{-3}$
to 5.77 $\times$ 10$^{-3}$ fm$^{-3}$. One can conclude
that the pairing effects are very small on these cells. For the
higher density region $\rho \sim$ 8.91 $\times$ 10$^{-3}$ fm$^{-3}$
to 4.75 $\times$ 10$^{-2}$ fm$^{-3}$, corresponding to
cells  $^{1350}$Sn, $^{1800}$Sn and  $^{1500}$Zr, 
there are differences between the density profiles 
due to the pairing effects. This is because the occupancy of the
8$s_{1/2}$ state is modified in HF-BCS calculations. Thus, the
difference between HF 
and HF-BCS neutron densities at r=0 is
around 10-11\% for the three cells above. Indeed, the occupancies of the
8$s_{1/2}$ neutron orbital are 0.91, 0.98 and 0.95 for cells
$^{1350}$Sn, $^{1800}$Sn and  $^{1500}$Zr in HF-BCS,
respectively, while this state is fully filled in HF calculations
of these cells. In spite of the cell $^{1500}$Zr, the HF-BCS neutron
densities of the cells $^{1350}$Sn, $^{1800}$Sn have an extended
``surface" before they reach the constant values corresponding to
the neutron gas. The neutron gas densities of these cells
are similar in both HF and HF-BCS approaches. For the
highest-density cell $^{982}$Ge with $\rho = $7.89 $\times$ 10$^{-2}$
fm$^{-3}$, the behaviour of the neutron density is slightly changed
due to the pairing effects. Although the D1S interaction
is used in the pairing channel, it cannot produce a constant
density around the outer edge of this cell. It seems that the cell
$^{982}$Ge most probably belongs to the deformed pasta phase.

\begin{figure*}[htb]
 \hspace*{0cm}\mbox{\epsfig{file=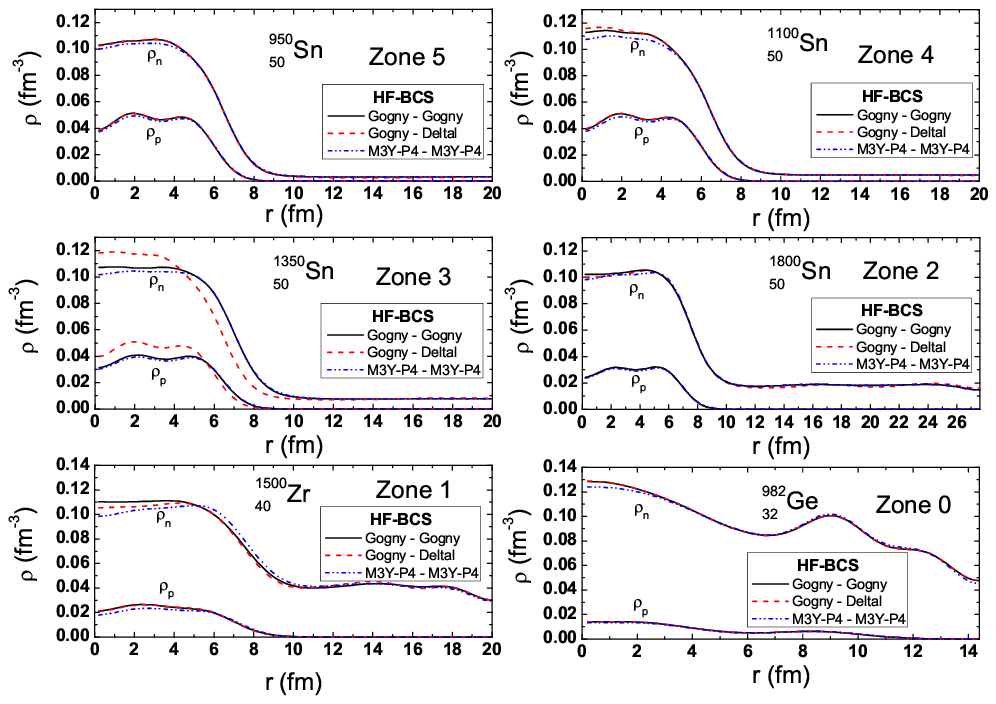,height=10cm}}
\caption{\small (Color online) The HF-BCS proton and neutron densities obtained
in the three cases indicated in the panels. The notation ``Gogny-Gogny" means the D1S Gogny
interaction is used in the mean field and pairing channels, and similar
notations for other cases. The calculations are done from zone
5 to zone 0. } \label{f23}
\end{figure*}

Since it is not yet well established what are the pairing properties of
neutron matter, we perform again the HF-BCS calculations for all
WS cells of the inner crust matter using three different pairing interactions, namely
 the D1S, M3Y and delta
interactions. 
The results of proton and
neutron density distributions obtained with these three pairing interactions
are shown in Fig. \ref{f23}.
Different combinations of mean fields and pairing interactions are used in this figure.
One observes 
that the
main features of the WS cells can be obtained with three kinds of
pairing interactions. At the high density region, except for the cell
$^{1350}$Sn where the pairing effects on the neutron
density are strong, the calculated proton and neutron densities are similar
for the other cells such as $^{1100}$Sn, $^{1500}$Zr,
$^{1800}$Sn and $^{982}$Ge. In the case of the $^{1350}$Sn cell, the
neutron density distribution of the nuclear cluster 
calculated with the zero-range pairing force
is higher than those
obtained with D1S or M3Y-P4 pairing interactions 
because the occupancy of the 8$s_{1/2}$ neutron
orbital corresponding to the delta-pairing interaction is larger than
those of the D1S and M3Y-P4 pairing interactions. In all
the cells we observe that the density distribution of the nuclear
clusters obtained with the M3Y-P4 interaction is slightly less extended  
than those obtained with the D1S interaction. This
effect may come from the differences of the ranges and the values of
pairing gaps in infinite matter corresponding to these two effective
interactions.

\subsection{Pairing fields}

Finally, we discuss briefly the BCS neutron pairing fields 
in the case of a delta
-pairing interaction. 
Rewriting Eq.~(\ref{th73}) as
\begin{equation}
V(\bm{r}_1-\bm{r}_2) = V_{eff}(\rho(r))\delta(\bm{r}_1-\bm{r}_2)~,
\end{equation}
the pairing field $\Delta(r)$ is a local function which can be expressed in terms
of the local pairing density $\kappa(r)$ as
\begin{equation}
\Delta(r)=V_{eff}(\rho(r))\kappa(r)~, \label{th75}
\end{equation}
where the BCS pairing density $\kappa(r)$  
is
\begin{equation}
\kappa(r)=\frac{1}{4\pi} \sum_i u_iv_i |\varphi_(r)|^2~.
\label{th76}
\end{equation}
Here, the factors $u_i$ and $v_i$ 
are the usual BCS amplitudes.

\begin{figure*}[htb]
\hspace*{1cm}
 \mbox{\epsfig{file=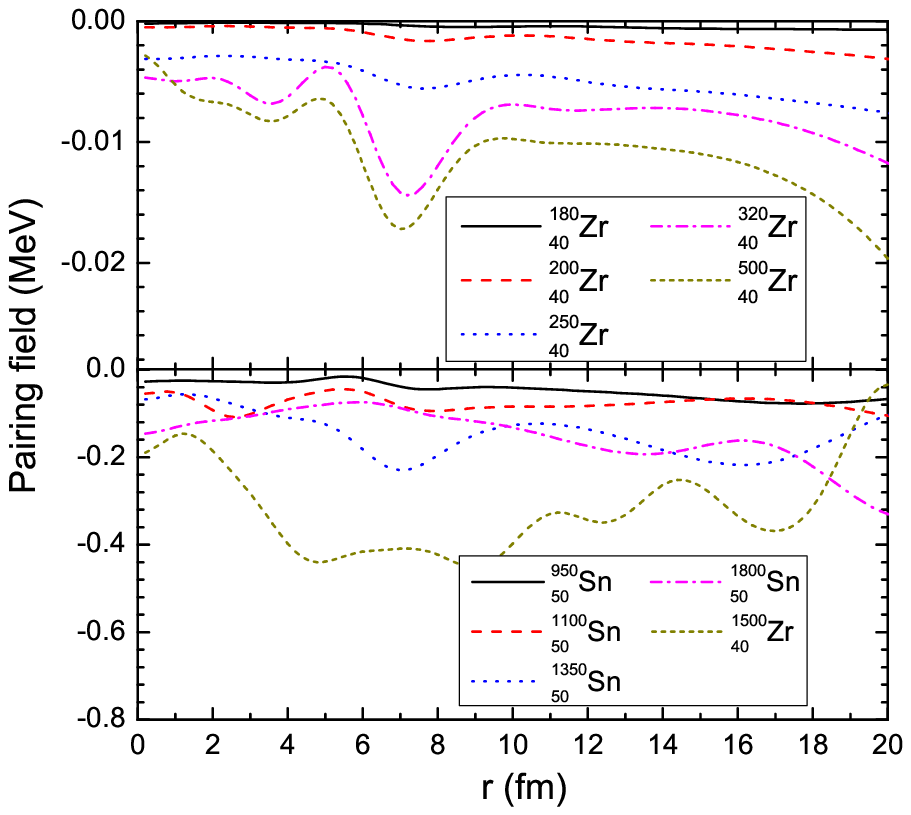,height=9cm}}
\caption{\small (Color online) The neutron pairing field $\Delta(r)$ of
Eq.(\ref{th75}) in zones 10 to 1, calculated with the zero range pairing
interaction and the D1S interaction in the particle-hole mean field}
\label{f24}
\end{figure*}

The results calculated for the WS cells from zone 10 to zone 1 (using D1S in
the particle-hole mean field) are shown
in Fig. \ref{f24}, except for zone 0 
which belongs to the deformed pasta phase. In the upper panel of
Fig.~\ref{f24}, one can see that the pairing field is becoming very small
almost everywhere in the cell. However, one can observe that in passing from
the low density region of the neutron gas towards the higher density region of
the cluster, the pairing field is increasing in the intermediate density
region of the cluster surface in the cells $^{320}$Zr and $^{500}$Zr. This
is a manifestation of the bell shape dependence of the pairing gap on
density. For the cells corresponding to high baryonic densities shown in
the lower panel of Fig.~\ref{f24} the slope of the pairing field is changing
very slowly when it is crossing the region between the nuclear cluster and
the uniform neutron gas. Using the same density-dependent delta force in the
pairing channel and the D1S force in the mean field channel,
the pairing fields of the cells $^{950}$Sn, $^{1500}$Zr and $^{1800}$Sn
are about three times smaller (in absolute value) than
those obtained by the HFB calculation using the SLy4 interaction in the mean
field (see Figs. 5 and 6 in Ref. \cite{Ni04}). We see that the pairing
correlations are reduced strongly for all the cells in the present work, due
to the use of finite range interactions in the mean field.

One can conclude that the behaviour of the pairing field in the inner crust
matter is rather complex. On the other hand the magnitude of the pairing
field inside the inner crust depends strongly on the scenario used for the
pairing properties of infinite neutron matter as shown in Refs. \cite{Ni04,
Ni04b}.

\section{Conclusions and Perspectives}

In this work, we have studied the properties of the WS cells in the inner
crust matter of neutron stars using 
finite-range,
density-dependent interactions in HF and HF-BCS approximations. Calculations
are performed for 11 representative WS cells by imposing Dirichlet-Neumann
boundary conditions 
at the edge of the cell. The study is done with the
D1S and M3Y-P4 effective interactions in both mean
field and pairing channels. For the pairing field we have also used a
density-dependent, zero-range force whose parameters
have been fixed to reproduce the pairing properties of infinite
neutron matter as predicted by microscopic calculations which take into account
polarization effects \cite{Ni04,Wa93}. For the HF-BCS calculations using the
zero-range pairing force, the HF mean field is calculated with the D1S interaction.

With the three different pairing interactions (D1S, M3Y-P4 and delta force), it
is found that the behaviour of the proton and neutron density distributions
are very similar 
in the low
density region, $\rho \sim$ 2.79 $\times$ 10$^{-4}$ fm$^{-3}$ to 
$\rho \sim$ 5.77 $\times$ 10$^{-3}$ fm$^{-3}$. One can conclude that the pairing effects
are very small in this region. In the higher density region, the pairing effects
can make an extended ``surface" for the neutron density distribution in the
cell before they reach a constant density around the outer edge of the cell.
It should be noted that pairing effects are weakened by the use of finite
range interactions in the particle-hole (mean-field) channel, compared to the use of
Skyrme interactions. However, one cannot obtain a constant neutron gas
density in the outer part of the $^{982}$Ge cell with finite-range density-dependent
interactions. This cell corresponds to the highest density $\rho$ = 7.89
$\times$ 10$^{-2}$ fm$^{-3}$, and it seems not to belong to the spherical
case as assumed in our study.

In all the cells we observe that the density
distribution of the nuclear cluster obtained with the case of M3Y-P4
interaction is slightly smaller than those obtained with the D1S
interaction. This effect may come from the difference of the ranges and the
values of pairing gaps in infinite matter given by the two kinds of
effective interactions. The largest effect on the densities comes from the
use of finite range interactions in the ``particle-hole'' mean field,
compared to the use of Skyrme interaction. This result shows that finite
range interactions should be considered in order to describe microscopically
Wigner-seitz cells.

\section*{Acknowledgements}
We wish to thank N. Sandulescu for
fruitful discussions. This research project has been supported, in part, by
Vietnam Natural Science Council and Vietnam Atomic Energy Commission and
also the ANR NExEN. H.S.T. acknowledges the financial
support from the Asia Link Programme CN/Asia-Link 008 (94791) and the Bourse
Eiffel program of the French Ministry of Foreign Affairs.

\appendix

\section{Hartree-Fock potentials with finite range interactions}

We give here some general expressions of HF potentials corresponding to Gogny-type forces.
They can be easily adapted to the case of the M3Y-P4 Nakada's interaction \cite{Na09} if one adopts a zero-range
approximation for the two-body spin-orbit component, as mentioned in subsec. II.A~.

The contributions of the central force
are separated into a Hartree (direct) and a Fock (exchange) potential:
\begin{eqnarray}
U^H_i(r_1)&=&\sum_j\sum^2_{\nu=1}\widehat{j_j}^2(
W_\nu+\frac{B_\nu}{2} - H_\nu\delta_{q_iq_j}-
\frac{M_\nu}{2}\delta_{q_iq_j} ) \nonumber \\
&\times &\int u^2_j(r_2) v^\nu_{0}(r_1,r_2)
r^2_2 dr_2~, \label{th26}
\end{eqnarray}


\begin{eqnarray}
U^F_i(r_1,r_2)&=&\sum_{jL}\sum^2_{\nu=1}\hat{j_j}^2
u_j(r_1)u_j(r_2) v^\nu_L(r_1,r_2) \nonumber \\
& \times &\3j{l_i}{l_j}{L}{0}{0}{0}^2  
\left[ \hat{l_i}^2\hat{l_j}^2\6j{l_i}{j_i}{1/2}{j_j}{l_j}{L}^2\right. \nonumber \\
& \times &\left.(W_\nu\delta_{q_iq_j}-H_\nu)
+B_\nu\delta_{q_iq_j}-M_\nu \right]~,\label{th27}
\end{eqnarray}
with $|l_i-l_j|\leq L \leq (l_i+l_j)$, and we use the notation $\hat{j}=(2j+1)^{1/2}$.

The density-dependent component of the force is similar to the zero-range density-dependent term of
Skyrme interactions \cite{Vau72,Ch98}, and we recall for completeness the corresponding contribution
(direct plus exchange) to the mean field: 

\begin{eqnarray}
U^{DD}_q(r)=\frac{t_3}{24}
\lbrace
(2+x_3)(2+\alpha)\rho^{\alpha+1}(r) - (2x_3+1) \nonumber \\
 \times [2\rho^\alpha(r)\rho_q(r)+\alpha\rho^{\alpha-1}(r)(\rho^2_p(r)+\rho^2_n(r)) ]  \rbrace
~,\nonumber \\
\label{th28}
\end{eqnarray}
where $t_3$, $x_3$ and $\alpha$ are the parameters of the density-dependent force,
$\rho$ is the total nucleon density and $q$ stands for  protons or neutrons.

The direct plus exchange spin-orbit mean field is \cite{Ch98}:
\begin{eqnarray}
V^{LS}_q(r &)&=W_0 \left \{
\frac{1}{r}\frac{d}{dr}(\rho(r)+\rho_q(r) )\bm{l}.\bm{s}\right.\nonumber \\
&-&\left. \left [\frac{1}{r}J(r) + J'(r) + \frac{1}{r}J_q (r) +
J'_q(r) \right ] \right \} \nonumber \\
&=& W^{LS0}_q(r)\bm{l}.\bm{s}+W^{LS1}_q(r)~, \label{th29}
\end{eqnarray}
where $J(r)$ is the spin density, and $J'= \frac{dJ}{dr}$.

Finally, the direct and exchange Coulomb mean fields are:
\begin{eqnarray}
V^{DC}(r_1)&=&e^2\sum_{j \in \text{protons}} \hat{j_j}^2 \int u^2_j(r_2)v^C_0(r_1,r_2)r_2^2
dr_2 \nonumber \\
&=& e^2 \int \rho_p(r_2)v^C_0(r_1,r_2)r_2^2 dr_2~, \label{th30}
\end{eqnarray}
\begin{eqnarray}
V^{EC}_i(r_1,r_2)=e^2\sum_{jL}\delta_{q_i,-1/2}\delta_{q_i,q_j}\hat{l_i}^2\hat{l_j}^2\hat{j_j}^2u_j(r_1)u_j(r_2)
\nonumber \\ \times v^C_L(r_1,r_2)  \3j{l_i}{l_j}{L}{0}{0}{0}^2
\6j{l_i}{j_i}{1/2}{j_j}{l_j}{L}^2~,\label{th31}
\end{eqnarray}
where $v^C_L(r_1,r_2)=r^L_</r^{L+1}_>$ are the multipoles of $1/\vert {\bf r}_1-\vert {\bf r}_2 \vert$.

\section{Pairing matrix elements}
Here, we give the main expressions for calculating the pairing fields in the BCS approximation.
The gap equation is \cite{Ri80}: 
\begin{equation}
\Delta_a=-\frac{1}{2}\sum_b (-1)^{l_a+l_b}\hat{j_a}^{-1}\hat{j_b}
G_0(aabb)
\frac{\Delta_b}{\sqrt{(\varepsilon_b-\lambda)^2+\Delta^2_b}}~,
\label{th203}
\end{equation}
where
\begin{equation}
G_0(aabb)=<aa|V_p(1,2)|bb>_{00} \label{th204} \end{equation}
is the $J=0$ particle-particle matrix element of the $V_p(1,2)$ pairing interaction.


\subsection{Particle-particle matrix elements with a finite-range interaction}
For any interaction of the D1S or M3Y-P4 type the general particle-particle matrix element
of the finite range part is:
\begin{eqnarray}
&&<ac|V_p(1,2)|bd>_{JM}\nonumber \\
&=& \int r^2_1dr_1 r^2_2dr_2\sum_{\nu
LSK}(-1)^{\mathcal{P}}\nonumber \\
&\times & A_\nu(S) v^\nu_L(r_1,r_2)  R_a(r_1)R_b(r_1)
R_c(r_2)R_d(r_2)\nonumber \\
&\times & \6j{j_d}{j_b}{J}{j_a}{j_c}{K}
<\mathcal{Y}_a\|T^{(SL)K}_{(1)}\|\mathcal{Y}_b> \nonumber \\
&\times &<\mathcal{Y}_c\|T^{(SL)K}_{(2)}
\|\mathcal{Y}_d>~.  \label{th115}
\end{eqnarray}


Here, the total phase is $\mathcal{P}=L+S+K+J+j_b+j_c$, $A_\nu(S)$ is given by 
Eq. (\ref{th39}),
$R_i(r)$ is the radial part of the single-particle wave function of Eq. (\ref{th19*}), and \\ 
$<\mathcal{Y}_i\|T^{(SL)K}_{(1)}\|\mathcal{Y}_j>$ is the reduced matrix element:
\begin{eqnarray}
\left < l_ij_i\|T^{(SL)J}\|l_jj_j \right > =
(-)^{l_i}\frac{\sqrt{2}}{\sqrt{4\pi}}\hat{S}\hat{L}\hat{J}\hat{l_i}\hat{l_j}\hat{j_i}\hat{j_j} \nonumber \\
\3j{l_i}{l_j}{L}{0}{0}{0}\9j{j_i}{j_j}{J}{l_i}{l_j}{L}{\frac{1}{2}}{\frac{1}{2}}{S}.
\label{th23}
\end{eqnarray}

For the pairing
matrix elements $<aa|V_p(1,2)|bb>_{00}$ we thus obtain:
\begin{eqnarray}
<aa|V_p(1,2)|bb>_{00}&=&\hat{j_a}^{-1}\hat{j_b}^{-1}\int r^2_1dr_1 r^2_2dr_2
\sum_{\nu LSK} A_\nu(S)\nonumber \\
&\times &v^\nu_L(r_1,r_2) R_a(r_1)R_b(r_1)
R_a(r_2)R_b(r_2) \nonumber \\
&\times &(-1)^{L+S} |<\mathcal{Y}_a\|T^{(SL)K}\|\mathcal{Y}_b>|^2~.
\label{th116}
\end{eqnarray}

The zero-range part of the Gogny or M3Y-P4 interactions does not contribute to the pairing matrix elements,
while the small contribution of the zero-range spin-orbit component to the pairing field is neglected.

 \subsection{Particle-particle matrix elements with a zero-range interaction}
The density-dependent delta interaction is taken of the
form:
\begin{equation}
V_p(1,2)=V_0 \left ( 1-\eta \left
(\frac{\rho(r_{12})}{\rho_0}\right )^\alpha \right ) \delta
(\bm{r_1}-\bm{r_2}) \label{102}
\end{equation}
The delta function is expanded on the basis of spherical harmonics:
\begin{equation}
\delta(\bm{r_1}-\bm{r_2})=\frac{\delta(r_1-r_2)}{r_1r_2}\sum_{L\mu}(-1)^\mu
Y^\mu_{L}(1)Y^{-\mu}_L(2) \label{103}
\end{equation}
The calculation of the coupled matrix element is similar to that of subsection B.1.
The result is: 
\begin{equation}
<aa|V_p(1,2)|bb>_{00}=\frac{1}{2}(-1)^{l_a+l_b}\frac{\hat{j_a}\hat{j_b}}{4\pi}
I_{aabb}~,
\end{equation}
where
\begin{equation}
I_{aabb}=V_0\int \left( 1-\eta \left( \frac{\rho(r)}{\rho_0}
\right )^\alpha \right ) R^2_a(r)R^2_b(r) d\bm{r}~.
\end{equation}

\end{document}